\documentclass[aps,prb,showpacs,twocolumn]{revtex4}
\usepackage{amsmath,amssymb,mathrsfs,latexsym,graphicx,psfrag}
\usepackage[varg]{txfonts}

\newcommand{\bs}[1]{\boldsymbol{#1}}

\newcommand{\ket}[1]{\left|#1\right\rangle}

\newcommand{\ev}[1]{\bigl\langle#1\bigr\rangle}

\newcommand{\sket}[1]{|#1\rangle}
\newcommand{\sbra}[1]{\langle #1 |}

\newcommand{\up}{\uparrow}
\newcommand{\dw}{\downarrow}

\newcommand{\bk}{\bs{k}}

\renewcommand{\section}[1]{}

\def\ie{\emph{i.e.},\ }

\def\ea{\emph{et al.}}

\def\etc{\emph{etc.}\ }

\begin{document}
\title{Evidence for site-centered stripes from magnetic excitations in
  CuO superconductors} \author{Martin Greiter and Holger Schmidt}
\affiliation{Institut f\"ur Theorie der Kondensierten Materie, KIT,
  Campus S\"ud, D 76128 Karlsruhe}
\pagestyle{plain} \date{\today}
\begin{abstract}
  The success of models of coupled two-leg spin ladders in describing
  the magnetic excitation spectrum of La$_{2-x}$Ba$_x$CuO$_4$ has been
  widely interpreted as evidence for bond-centered stripes.  Here, we
  determine the magnetic coupling induced by the charge stripes
  between bond- or site-centered spin stripes modeled by two- or
  three-leg ladders, respectively.  We find that only the
  site-centered models order.  We further report excellent agreement
  of a fully consistent analysis of coupled three-leg ladders using a
  spin wave theory of bond operators with the experiment.
\end{abstract}
\maketitle

\section{Introduction}
The mechanism of superconductivity in the presence of local
antiferromagnetism in copper oxides is considered an outstanding
problem in contemporary
physics~\cite{zaanen-06np138,orenstein-00s468}.  The materials are
described by mobile charge carriers (holes) doped into a
quasi-twodimensional spin 1/2 antiferromagnet~\cite{Zhang-88prb3759}.
Inelastic neutron scattering experiments have revealed a magnetic
resonance peak~\cite{fong-95prl316,bourges-00s1234} and, in some
compounds, periodic modulations in the spin and charge density
(stripes)~\cite{zaanen-89prb7391,tranquada-95n561,emery-99pnas8814,zaanen-01pmb1485,mook-02prl097004,kivelson-03rmp1201,berg-09njp115004}.
Tranquada \ea~\cite{tranquada-04n534} found that the magnetic
excitation spectrum of stripe ordered La$_{1.875}$Ba$_{0.125}$CuO$_4$
looks similar to disordered
YBa$_2$Cu$_3$O$_{6+x}$~\cite{bourges-00s1234} or
Bi$_2$Sr$_2$CaCu$_2$O$_{8+\delta}$~\cite{fauque-07prb214512}, and
observed that the data are consistent with bond-centered stripes
modeled by two-leg ladders.  This experiment, and its interpretation,
is considered of key importance for the field of high-temperature
superconductivity.  The reason these findings are so vigorously
discussed is that they may provide the decisive hint as to within
which framework these conceptionally simple yet inscrutable systems
are to be understood, and hence ultimately lead to a complete theory.

At present, there is no consensus with regard to such a framework, but
instead a fierce competition among different schools of thought.  One
of these
schools~\cite{emery-99pnas8814,zaanen-01pmb1485,kivelson-03rmp1201,berg-09njp115004}
attributes the unusual properties of the doped, two-dimensional
antiferromagnets to their propensity to form stripes, or their
proximity to a quantum critical point (QCP) at which stripe order sets
in.  The resulting picture is highly appealing.  Static stripes have
been observed~\cite{tranquada-95n561,mook-02prl097004} only in certain
compounds, most notably La$_{2-x}$Sr$_{x}$CuO$_4$ at a hole doping
concentration $x=\frac{1}{8}$, and are known to suppress
superconductivity.  On the other hand, the mere existence of stripes
would impose an effective one-dimensionality, and hence provide a
framework to formulate fractionally quantized excitations.  This
one-dimensionality would be roughly consistent with an enormous body
of experimental data on the cuprates, including the electron spectral
functions seen in angle-resolved photo emission spectroscopy (ARPES).
The charge carriers, the holons, would predominantly reside in the
charge stripes, as they could maximize their kinetic energy in these
antiferromagnetically disordered regions.  In the spin stripes, by
contrast, the antiferromagnetic exchange energy between the spins
would be maximized, at the price of infringing on the mobility of the
charge carriers.  Most importantly, the spin stripes would impose a
coupling between the charge stripes, which would yield an effective,
pairwise confinement between the low-energy spinon and holon
excitations residing predominantly in the charge stripes.  The
mechanism of confinement would be similar to that of coupled spin
chains or spin
ladders~\cite{dagotto-96s618,shelton-96prb8521,greiter02prb054505}.
The holes would be described by spinon-holon bound states, and the
dominant contribution to the magnetic response measured in Tranquada's
as well as all other neutron scattering experiments would come from
spinon-spinon bound states.

\begin{figure}[t]
  \begin{center}
    \includegraphics[width=0.82\columnwidth]{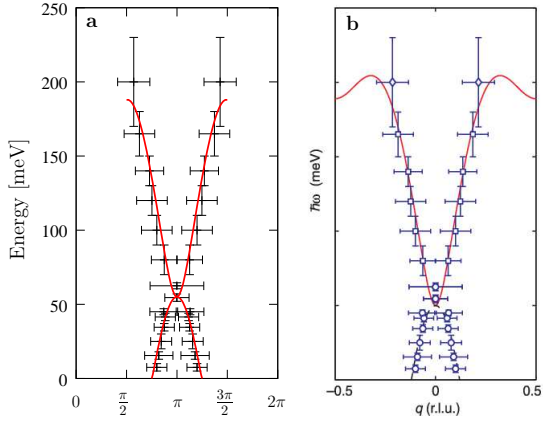}
  \end{center}
  \vspace{-12pt}
  \caption{(Color online)
    Comparison of dispersions resulting from different theories with
    experimental data.  {\bf a}, Superpositions of cuts along $(k_x,\pi)$
    and $(\pi,k_y)$ for the lowest mode of the SWT with
    $J=140\,\text{meV}$ and $J'=0.07\,J$ for site-centered stripes
    described in the text (red) superimposed with the experimental
    data obtained by inelastic neutron
    scattering~\cite{tranquada-04n534} (black).  {\bf b}, The neutron data,
    with a triplon dispersion of a two-leg ladder superimposed (red
    line), reproduced from Tranquada \ea~\cite{tranquada-04n534} (Reprinted by permission
from Macmillian Publishers Ltd: Nature 429: 534-538, \copyright 2005).}
  \label{fig:tddtr}
\end{figure}

\begin{figure}[tbh]
  \begin{center}
    \includegraphics[width=0.75\columnwidth]{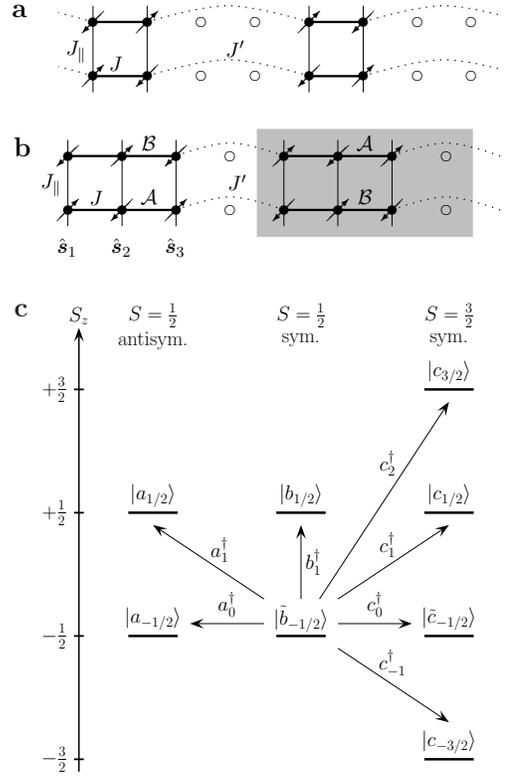}
  \end{center}
  \vspace{-8pt}
  \caption{%
    Models of spin stripes: {\bf a}, Bond-centered stripes modeled by
    two-leg ladders and {\bf b}, Site-centered stripes modeled by
    three-leg ladders. 
    {\bf c}, Definition of bosonic creation
    operators in the eight-dimensional Hilbert space spanned by a rung
    on sublattice $\cal A$ of the ladders shown in {\bf b}.}
  \label{fig:2leg_3leg}
\end{figure}

The similarity of the ``hour-glass'' spectrum shown in Fig.~4b of
Tranquada \ea~\cite{tranquada-04n534} (which is reproduced for
comparison in Fig.~\ref{fig:tddtr}B) and the ``elephants trousers''
observed by Bourges et al.~\cite{bourges-00s1234} (see Fig.~3 of their
manuscript) provides the most striking evidence in favor of the
picture advocated by this school, which attributes the anomalous
properties of generic, disordered CuO superconductors to the formation
of dynamic (rather than static) stripes, which fluctuate on time
scales slow compared to the energy scales of most experimental probes.
This picture received additional support by Xu \ea~\cite{Xu-07prb014508}, 
who observed that the magnetic response of
La$_{1.875}$Ba$_{0.125}$CuO$_4$ at higher energies is independent of
temperature, while the stripe order melts at about $T_{\rm
  st}\sim54{\rm K}$.  Measurements on `untwinned' samples of
YBa$_2$Cu$_3$O$_{6.6}$, where one would expect the dynamical stripes
to orient themselves along one of the axis, however, exhibit a strong
anisotropy in the response only at energies below the
resonance~\cite{hinkov-04n650}, while the response is fourfold
rotationally symmetric at higher energies~\cite{hinkov-07np780}.  This
may indicate that the formation of stripe correlations, be it static
or dynamic, is a low energy phenomenon, while the high energy response
probes itinerant antiferromagnets at length scales on which the
stripes are essentially invisible.
Another extremely appealing feature of the experiment by Tranquada
\ea~\cite{tranquada-04n534} is that it immediately suggests a model of
ferromagnetically coupled two-leg ladders, as the upper part of the
measured spectrum agrees strikingly well with the triplon (or
spinon-spinon bound state) mode of isolated two-leg ladders (see
Fig.~\ref{fig:tddtr}B).  The experiment hence appears to point
bond-centered rather than site-centered stripes (\ie stripes as depicted
in Fig.~\ref{fig:2leg_3leg}A rather than Fig.~\ref{fig:2leg_3leg}B),
and thereby to resolve a long outstanding issue within the field of
static stripes.

This interpretation received support by theoretical
studies~\cite{vojta-04prl127002,uhrig-04prl267003,anisimov-04prb172501}.
Vojta \ea~\cite{vojta-04prl127002} used a bond operator
formalism~\cite{sachdev-90prb9323} to study a spin-only model of
coupled two-leg ladders (as depicted in Fig.~\ref{fig:2leg_3leg}A)
with $J_{\parallel}=J$, took into account a bond-boson
renormalization 
of $J$, and assumed a value $J'$ for the ferromagnetic coupling
between the ladders which is large enough to close the spin gap of the
ladders, \ie to induce long range magnetic order.  Within their
approximations, a value of $J'=-0.06J$ is sufficient.  This value is
not consistent with previous
studies~\cite{gopalan-94prb8901,tworzydlo-99prb115,dalosto-00prb928},
but as no method to calculate or even estimate the true $J'$ induced
by charge stripes was available, it did not seem a problem at the
time.  The spectrum they obtained agrees well with experimental data
measured by Tranquada \ea~\cite{tranquada-04n534}, and hence appeared
to justify their assumptions \emph{a posteriori}.  They concluded in
favor of bond-centered stripes.
This conclusion was independently strengthened by Uhrig
\ea~\cite{uhrig-04prl267003}, who used the method of continuous unitary
transformations to study a model of ferromagnetically coupled two-leg
ladders, and observed that the critical value of $J_{\rm c}'$ can be
significantly reduced if a cyclic exchange term $J_{\rm cyc}$ on the
ladders is included~\cite{nunner-02prb180404}.  They likewise
fine-tuned $J'$ to the QCP where the gap closes and long-range
magnetic order ensues, and reported good agreement with experiment.
%
%
Seibold \ea~\cite{seibold-05prl107006,seibold-06prb144515} calculated
the magnetic response for a range of dopings within the time-dependent
Gutzwiller approximation, and found good agreement 
with the measured data for both bond- and site-centered stripe models.

\section{Numerical evaluation of inter-ladder couplings}
The first question we wish to address here is whether the key
assumption that $J'$ is large enough to induce order is valid.  There
exist several estimates for the critical value $J_{\rm c}'$ required
if the coupling between isotropic ladders is antiferromagnetic.
Gopalan \ea~\cite{gopalan-94prb8901} find $J_{\rm c}'\approx 0.25J$ in
a simple mean-field treatment of bond-bosons.  Quantum Monte Carlo
(QMC) calculations by Tworzyd\l{}o \ea~\cite{tworzydlo-99prb115} yield
$J_{\rm c}'=0.30(2)J$, a value subsequently confirmed by Dalosto
\ea~\cite{dalosto-00prb928}.  We have redone the mean-field calculation
of Gopalan \ea~\cite{gopalan-94prb8901} for ferromagnetic (FM)
couplings $J_{\rm c}'<0$, and find that within this approximation, the
absolute value of $J_{\rm c}'$ is independent of the sign of the
coupling.  QMC calculations by Dalosto \ea~\cite{dalosto-00prb928},
however, indicate that the true value is at least $J_{\rm c}=-0.4J$
(see Fig.\ 6b of their article).  The physical reason why a
significant coupling between the ladders is required to induce
magnetic order is that the individual two-leg ladders possess a gap of
order $\Delta\approx J/2$.  As a cyclic exchange term $J_{\rm
  cyc}\approx 0.25J$ reduces this gap by a factor of
two~\cite{nunner-02prb180404}, we would expect that $J_{\rm c}'$ would
likewise be reduced by a factor of two.  We hence conclude that a FM
coupling of at least somewhere between $J_{\rm c}'=-0.2J$ and $-0.4J$
is required, depending on the strength of a possible cyclic exchange
term.

The explicit calculation of the ferromagnetic coupling induced by the
charge stripes between the spin stripes we describe
below, however, yields $J'=-0.05J$.  The coupling
is hence insufficient to induce order in a model of bond centered stripes. 

For a model of site-centered stripes described by antiferromagnetically
coupled three-leg ladders, as shown in Fig.~\ref{fig:2leg_3leg}B, the
critical coupling required for long range ordering to set in is by
contrast $J_{\rm c}'=0$.  The reason is simply that there is no need
to close a gap, as the three-leg ladders are individually
gapless~\cite{dagotto-96s618}. 
A conventional 
spin wave analysis for such a spin-only model of three- and four-leg
ladders was performed by Yao \ea~\cite{yao-06prl017003}, who found that
their approximation agrees reasonable well with the experimental data
if they take $J'=0.05J$ and $J'=-0.09J$ for coupled three- and
four-leg ladders, respectively.  The calculation we present in the
following, however, singles out $J'=0.07J$ for the antiferromagnetic
coupling between spin stripes modeled by three leg ladders.

\begin{figure}[t]
  \begin{center}
    \includegraphics[width=0.75\columnwidth]{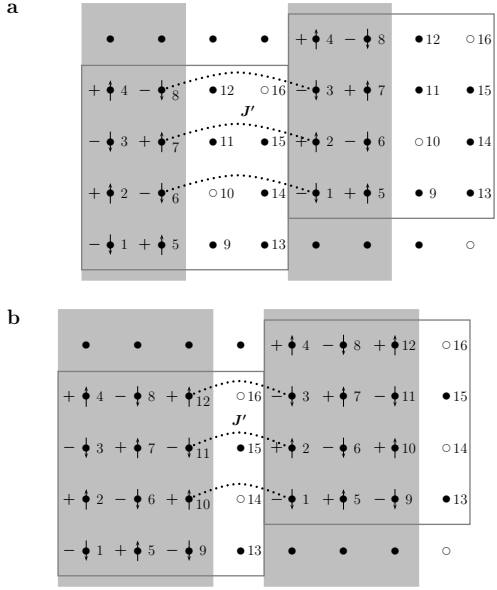}
  \end{center}
  \vspace{-8pt}
  \caption{%
    Finite size geometries with unfrustrated periodic boundaries for
    {\bf a}, bond- and {\bf b}, site-centered stripe models.  The spin
    stripes are localized by a staggered magnetic field $B$ as
    indicated by the signs in the grey shaded areas.}
  \label{fig:jperp}
\end{figure}

To determine the effective coupling $J'$ between the two- and
three-leg ladders representing bond- or site-centered stripes,
respectively, we have exactly diagonalized 16 site clusters of
itinerant spin 1/2 antiferromagnets described by the (nearest-neighbor)
$t$--$J$ model~\cite{Zhang-88prb3759} with $J=0.4t$, two holes, and
periodic boundary conditions (PBCs), in which the stripes are
localized through a staggered magnetic field $B$ in the grey shaded
areas shown in Figs.~\ref{fig:jperp}A and \ref{fig:jperp}B.  We then
compare the ground state energies we obtain for clusters with the
unfrustrated PBCs shown in Fig.~\ref{fig:jperp} with the ground state
energies we obtain for clusters with frustrated PBCs, in which the
16-site unit cells shown on the right are shifted by one lattice
spacing to the top, such that sites 15 and 1, 16 and 2, \etc are
nearest neighbors.  We then consider spin-only Heisenberg models of
two- and three-leg ladders (consisting of only the sites in the shaded
areas in Figs.~\ref{fig:jperp}A and \ref{fig:jperp}B) subject to the
same staggered field $B$ and couple them ferromagnetically or
antiferromagnetically by $J'$, respectively, as indicated.  We again
compare the ground state energies for unfrustrated PBCs, where $J'$
couples sites 6 and 1, 7 and 2, \etc for the two leg ladders shown in
Fig.~\ref{fig:jperp}A, with frustrated PBCs, where $J'$ couples sites
7 and 1, 8 and 2, \etc \ Finally, we determine $J'$ such that the
difference in the ground state energies between frustrated and
unfrustrated PBCs in the $t$--$J$ clusters matches this difference in
the spin-only ladder models.

\begin{figure}[t]
  \vspace{-8pt}
  \begin{center}
    \includegraphics[width=0.9\linewidth]{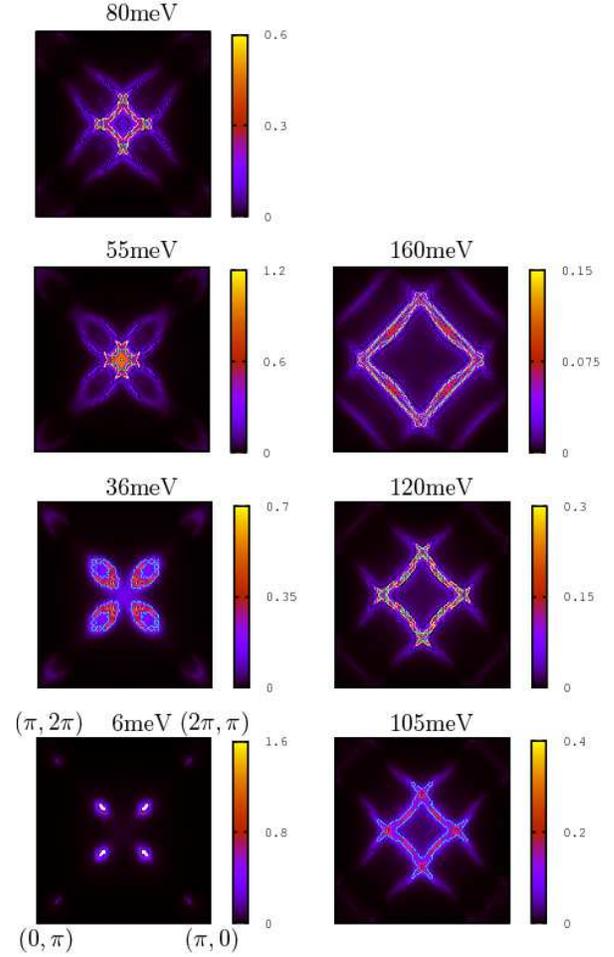}
  \end{center}
  \vspace{-8pt}
  \caption{(Color online)
    Constant energy slices of the neutron scattering intensity
    $\chi^{+-}(\bk,\omega)$ for the lowest mode of the SWT with
    $J=140\,\text{meV}$ and $J'=0.07\,J$.  We have replaced the
    $\delta$-functions in frequency by Lorentzians with half-width
    $\Delta=0.05\,J$ and averaged over both stripe orientations (\ie
    horizontal and vertical).}
  \label{fig:neutron}
\end{figure}

With $B=0.225J$ and $B=0.170J$ for the bond- or site-centered stripe
models we obtain $J'=-0.051J$ and $J'=0.071J$, respectively.  The
values for $B$ are chosen such that the magnetic localization energy
$E_{\text{mag.}}=-B\sum_i (-1)^i S^z_i$ is equal for both types of
stripes, and such that the mean value of the staggered magnetization
$\ev{-S^z_1+S^z_5}$ in the $t$--$J$ cluster for the site-centered
stripe matches the value we obtain in the 
spin wave theory (SWT) of three-leg ladders described below.
We estimate the error resulting from finite size corrections and the
use of a nearest-neighbor $t$--$J$ model to be of order $\pm0.01J$.

\section{Spin wave theory of three site rungs}
In the remainder of this Rapid Communication, we will show, albeit not
in full detail, that a fully consistent SWT of bond operators
representing the eight-dimensional Hilbert spaces on each rung of the
three-leg ladders agrees perfectly with the experimental data if and
only if the correct, calculated value $J'=0.07J$ is used for the
coupling.  This discussion will complete our argument showing that the
experiment by Tranquada \ea~\cite{tranquada-04n534} provides evidence
for site-centered, and not bond-centered, stripes.

To begin with, consider a single rung on sublattice $\cal A$ with
three spins $\bs{s}_1$,$\bs{s}_2$, and $\bs{s}_3$, as indicated in
Fig.~\ref{fig:2leg_3leg}B.  Diagonalization of the Hamiltonian $\hat
H^{\cal A} = J(\hat{\bs{s}}_1\hat{\bs{s}}_2 +
\hat{\bs{s}}_2\hat{\bs{s}}_3)$ yields a spin doublet
$\left\{\ket{b_{-1/2}},\ket{b_{1/2}}\right\}$ with energy $E=-J$ and a
quadruplet
$\left\{\ket{c_{-3/2}},\ket{c_{-1/2}},\ket{c_{1/2}},\ket{c_{3/2}}\right\}$
with $E=\frac{J}{2}$, both of which are symmetric under mirror
reflection (interchange of sites 1 and 3), and another doublet
$\left\{\ket{a_{-1/2}},\ket{a_{1/2}}\right\}$ with $E=0$, which is
antisymmetric under this mirror reflection.  The subscripts in the
kets label the eigenvalues of $\hat{s}^z_{\rm tot}$.  We define a
fiducial state $\sket{\tilde{b}_{-1/2}}$ for sublattice $\cal A$ via
$\sket{\tilde{b}_{-1/2}}
\equiv\sket{b_{-1/2}}\cos\phi+\sket{c_{-1/2}}\sin\phi $, which
interpolates between the $s^{z}_{\rm tot}=-\frac{1}{2}$ quantum ground
state $\sket{b_{-1/2}}$ of the isolated rung for $\phi=0$ and the
classically N\'eel ordered state $|\!\!\!\dw\up\dw\rangle$ for
$\phi=\arctan(\frac{1}{\sqrt{2}})=0.6155$.  We then introduce bosonic
operators $a_0^{\dagger}\equiv\sket{a_{-1/2}}\sbra{\tilde{b}_{-1/2}}$
\emph{etc.}, as indicated in Fig.~\ref{fig:2leg_3leg}C.  Written in
terms of these operators, the Hamiltonian $H^{\cal A}$ of a single
rung will contain a linear term
$\frac{3}{4}\bigl(c_0^{\dagger}+c_0^{\phantom{\dagger}}\bigr)\sin2\phi$.
Finally, we introduce a similar formalism with operators
$A_0^{\dagger}\equiv\sket{a_{1/2}}\sbra{\tilde{b}_{1/2}}$ \etc and a
fiducial state $\sket{\tilde{b}_{1/2}}$ with $S^{z}_{\rm
  tot}=+\frac{1}{2}$ for the rungs on sublattice $\cal B$, and express
all the spin operators on all the individual sites in terms of the
bosonic operators.

When we couple the rungs with intra-ladder couplings $J_{\parallel}=J$
and inter-ladder couplings $J'$ (see Fig.~\ref{fig:2leg_3leg}B), we
keep terms up to quadratic order in the bosonic operators, and
adjust the angle $\phi$ introduced above such that the terms linear in
the operators vanish.  This yields $\phi=0.341$ for $J'=0.07J$.  We
introduce Fourier transforms of our bosonic operators into momentum
space using the unit cell indicated by the grey area in
Fig.~\ref{fig:2leg_3leg}B, and rewrite the total Hamiltonian in terms
of those.  As the linear SWT preserves both the mirror reflection
symmetry and the total spin quantum number $S^z$, the Hamiltonian
separates into several terms, $\hat H = \tilde E_0+\hat H_{a0} + \hat
H_{c0} + \hat H_{a1} + \hat H_{c2} + \hat H_{b1,c1,c-1}$, where
$\tilde E_0$ is a contribution to the ground state energy, $\hat
H_{a0}$ contains only the operators $a_0^{\dagger}$,
$a_0^{\phantom{\dagger}}$, $A_{0}^{\dagger}$, and
$A_{0}^{\phantom{\dagger}}$, and so on.  The low-energy spectrum we
are interested in is exclusively contained in 
\begin{equation}\label{eq:hs}\nonumber
\hat H_{b1,c1,c-1} = \sum_{\bk}\Big( 
\hat\Psi_{\bk}^{\dagger}H^{\phantom{\dagger}}_{\bk}\hat\Psi^{\phantom{\dagger}}_{\bk}
- \frac{1}{2}\text{tr}(H_{\bk})\Big),
\end{equation}
where $\hat\Psi^{\phantom{\dagger}}_{\bk} \equiv \bigl(
B_{-1,\bk}^{\phantom{\dagger}}, b^{\dagger}_{1,\bk},
C_{-1,\bk}^{\phantom{\dagger}}, c^{\dagger}_{1,\bk},
C^{\dagger}_{1,-\bk}, c_{-1,-\bk}^{\phantom{\dagger}}\bigr)^{\rm T}$
and $H_{\bk}$ is a $6\times 6$ matrix.  We diagonalize $\hat H_{b1,c1,c-1}$ via a
six-dimensional Bogoliubov transformation at each point in $\bk$
space.

Cuts of the dispersion of the lowest mode we find are shown in
Fig.~\ref{fig:tddtr}A.  They agree strikingly well with the
superimposed experimental data by Tranquada
\ea~\cite{tranquada-04n534}.  Constant energy slices of the neutron
scattering intensity $\chi^{+-}(\bk,\omega)$ of this mode, which may
be compared to the data reported in Fig.\ 2 of Tranquada
\ea~\cite{tranquada-04n534}, are shown in Fig.~\ref{fig:neutron}.  The
other two modes of $H_{b1,c1,c-1}$, as well as all modes from the
other terms in $\hat H$, are only weakly dispersing and start at
energies of order $2J$, \ie in a regime where we would no longer
expect our spin wave theory to yield reliable results.

Finally, let us remark that for $J'\lesssim 0.5 J$, the saddle point
energy in our model is given to a highly accurate approximation by
$\omega(\pi,\pi)\approx 1.47 \sqrt{J'J}$.  
Since we expect $J'$
to decrease with decreasing doping, we would expect a similar doping
dependence for the saddle point
energy~\cite{seibold-05prl107006,seibold-06prb144515}.

\section{Conclusion}
In conclusion, we have shown that models of coupled two-leg ladders
describing bond-centered stripes cannot explain the magnetic spectrum
of La$_{2-x}$Ba$_x$CuO$_4$~\cite{tranquada-04n534} as the coupling
induced by the charge stripes between the ladders is insufficient to
induce long-range magnetic order.  We have further shown that a model
of coupled three-leg ladders describing site-centered stripes accounts
accurately for the experimental data.  The experiment hence
provides evidence for site-centered, and not bond-centered, stripes.

We wish to thank M.~Vojta, T.~Ulbricht and P.~W\"olfle
for numerous discussions.  This work was supported by the German
Research Foundation under grant FOR 960.


\end{document}